%% file: paper.tex
\documentclass[twoside,fleqn]{article}

\usepackage{espcrc2,latexsym,wasysym,bbm,epsfig,rotating,here,citesort}

\begin{document}

\title{
\vspace{-0.6cm}
{\sf \normalsize \rightline{HD-THEP-00-52}}
{\Large Inelastic diffractive proton-proton scattering in\\ nonperturbative QCD}}
\author{O. Nachtmann and T. Paulus\thanks{Poster presented at the Workshop ``Diffraction 2000'', Cetraro, Italy, September 2000.\newline\newline
Supported by the German Bundesministerium f\"ur Bildung und Forschung (BMBF), Contract no. 05 HT 9VHA3 and by Deutsche Forschungsgemeinschaft (DFG), Grant no. GRK 216/1/00}
\address{
       Institut f\"ur Theoretische Physik, Universit\"at Heidelberg, \\
       Philosophenweg 16, 69120 Heidelberg, Germany} 
       }

\begin{abstract}
We examine diffractive proton-proton scattering $p \: p \to p \: X$.
Using a functional integral approach we derive the scattering amplitudes, which are governed by the expectation value of lightlike Wegner-Wilson loops.
This expectation value is then evaluated using a cumulant expansion and the model of the stochastic vacuum.
From the scattering amplitudes we calculate total and differential cross sections for high centre of mass energy and small momentum transfer and compare with experiments.
\end{abstract}

\maketitle

\input{intro}
\input{scatamp}
\input{eval}
\input{results}
\input{conclusion}
\section*{Acknowledgements}
We are grateful to E.R. Berger, H.G. Dosch, G. Kulzinger, M. R\"uter and S. Weinstock for many fruitful discussions.

\input{biblio}
\end{document}

%% file: intro.tex
\section{Introduction}

In this article we study diffractive proton-proton scattering $p \: p \to p \: X$ at high centre of mass (c.m.) energies $\sqrt{s} \apprge 20 \mbox{ GeV}$ and small momentum transfer squared $|t| \apprle 0.5 \mbox{ GeV}^2$.
The low momentum transfer implies that one has to apply nonperturbative methods to investigate these processes.

A description of soft hadronic high energy reactions, starting from a microscopic level, was developed in~\cite{LN} where in the case of an abelian gluon model the pomeron properties were related to nonperturbative aspects of the vacuum like the gluon condensate introduced by Shifman, Vainshtein and Zakharov~\cite{SVZ}. These methods were generalized to QCD in~\cite{Nachtmann}. The quantity governing the scattering amplitude was found to be a correlation function of two lightlike Wegner-Wilson loops~\cite{DFK}. These correlation functions are evaluated using the model of the stochastic vacuum (MSV)~\cite{MSV} after an analytic continuation from Euclidian space to Minkowski space~\cite{MSVmink}.

In this paper we will apply the model to inelastic diffractive proton-proton scattering. The hadronic scattering amplitude as derived in~\cite{Nachtmann,DFK,Nachtmannreport,DFK2} is given in section~\ref{scat_sect}, the result for the loop-loop correlation function is shown in section~\ref{eval_sect}. The numerical results are presented in section~\ref{result_sect} and compared to experimental data.

%% file: scatamp.tex
\section{The scattering amplitudes} \label{scat_sect}

In this section we present our basic formulae for hadron-hadron scattering. Consider the reaction 
\begin{equation}\label{hh_react}
h_1(P_1) + h_2(P_2) \to h_1(P_3) + X(P_4),
\end{equation}
where $h_1$ and $h_2$ are hadrons, $X$ is again $h_2$ or a diffractive excitation of $h_2$.
The hadrons $h_1, \, h_2$ are modeled as $q \bar q$ and quark-diquark wave packets for mesons and baryons respectively. For the wave functions we have chosen a Bauer-Stech-Wirbel type ansatz~\cite{BWS}.
The diffractive final state $X$ is modeled by a $q \bar q$-pair (or quark-diquark pair) in a colour singlet state. For the description of the quark and antiquark (diquark), we use free plane waves. Integration over all allowed values in phase space and the closure relation then yield all possible diffractive final states $X$, where the case of elastic scattering is also included.

In the framework of the model presented  in~\cite{DFK,Nachtmannreport} we obtain the scattering amplitude for reaction~(\ref{hh_react}) as
\begin{eqnarray} \label{scat_amp}
  \lefteqn{\mathcal{T}_{fi} = (2is)(2 \pi) \int_0^\infty \mbox{d}b_T \, b_T J_0 (\sqrt{-t} b_T) \hat J_{\mbox{\scriptsize{diff}}}.}
\end{eqnarray}
Here $J_0$ is the Bessel function of zeroth degree and $\hat J_{\mbox{\scriptsize{diff}}}$ is the diffractive profile function for which we obtain
\begin{eqnarray}
  \lefteqn{\hat J_{\mbox{\scriptsize{diff}}}({\bf b}_T,z^\prime) = - \int \mbox{d}^2x_T \, \mbox{d}^2y_T \int_0^1 \mbox{d}z \; w_{31}({\bf x}_T,z)} \nonumber \\
 & &\sqrt{2 \pi} \sqrt{2 z^\prime (1-z^\prime)} \, e^{-i {\bf \Delta}_{4T} \cdot {\bf y}_T} \, \varphi_2({\bf y}_T,z^\prime) \nonumber \\ 
 & &\Bigl \langle \mathcal{W}_+(\frac{1}{2}{\bf b}_T+(\frac{1}{2}-z){\bf x}_T,{\bf x}_T) \nonumber \\ 
 & &\;\; \mathcal{W}_-(-\frac{1}{2}{\bf b}_T+(\frac{1}{2}-z^\prime){\bf y}_T,{\bf y}_T)-\mathbbm{1}\Bigr \rangle_G, \label{J_wave}
\end{eqnarray}
where $w_{31}({\bf x}_T,z)$ denotes the profile function for the overlap between initial and final state of the hadron $h_1$ for fixed ${\bf x}_T$ and $z$. $\varphi_2({\bf y}_T,z^\prime)$ defines the initial state wave function of $h_2$.
Using the ansatz of~\cite{BWS} we have
\begin{eqnarray}
  \varphi_i({\bf x}_T,z) &=& \frac{1}{\sqrt{2 \pi S_{h_i}^2 I_{h_i}}} \, e^{-{\bf x}_T^2/4 S_{h_i}^2} \nonumber \\ 
  & &\sqrt{2 z (1-z)} e^{-(z-\frac{1}{2})^2/4 z_{h_i}^2}, \label{wave_funct} \\
  w_{ij}({\bf x}_T,z) &=& \varphi_i({\bf x}_T,z) \, \varphi_j({\bf x}_T,z).\label{profile_funct}
\end{eqnarray}
Here $I_{h_i}$ is a normalization factor.
The lightlike Wegner-Wilson loops $\mathcal{W}_\pm$ are given by
\begin{equation} \label{loop}
  \mathcal{W}_\pm:=\frac{1}{3} \,\mbox{tr P}\exp{(-ig \int_{C_\pm} \mbox{d}x^\mu G^a_\mu(x) \frac{\lambda^a}{2})},
\end{equation}
where P denotes path ordering and $C_\pm$ is the curve consisting of two lightlike worldlines for the quark and the antiquark (diquark) and connecting pieces at $\pm \infty$. ${\bf x}_T$ and ${\bf y}_T$ define the extension and orientation in transverse position space of the two loops representing the two hadrons $h_1$ and $h_2$ respectively, $z$ ($z^\prime$) parametrizes the fraction of the longitudinal momentum of hadron $h_1$ ($h_2$) carried by the quark. The impact parameter is given by ${\bf b}_T$.

The symbol $\langle\ldots\rangle_G$ denotes the functional integration which correlates the two loops. In~(\ref{J_wave}) the loop-loop correlation function is multiplied with the profile function $w_{31}$, the incoming wave function $\varphi_2$ and then integrated over all extensions and orientations of the loops in transverse space as well as over the longitudinal momentum fraction $z$ of hadron $h_1$, which is not diffractively excited. 
The hadronic scattering amplitude is then given by a Fourier-Bessel transform of this expression with respect to the impact parameter ${\bf b}_T$.

%% file: eval.tex
\section{Evaluation of the scattering amplitudes}\label{eval_sect}

Now we perform the functional integral in~(\ref{J_wave}) making use of the MSV. A detailed presentation of the MSV can be found in~\cite{DFK,MSV,MSVmink}, where both the original formulation in Euclidian space-time and the analytic continuation to Minkowski space-time are discussed.\\
The first step is to make a cumulant expansion~\cite{Nachtmannreport} for the loop-loop correlation function.
Proceeding as explained in~\cite{Berger} and assuming $|\chi| \ll 1$ we find
\begin{equation}\label{chi2_result}
  \left\langle \mathcal{W}_+ \mathcal{W}_- - \mathbbm{1}\right\rangle_G=-\frac{1}{9} \, \chi^2,
\end{equation}
which is the result of the traditional expansion method~\cite{DFK}.
Here we have defined
\begin{eqnarray} \label{chi}
   \lefteqn{\chi({\bf b}_T,{\bf x}_T,{\bf y}_T,z,z^\prime) = \frac{G_2 \pi^2}{24} \Bigl\{ I({\bf r}_{xq},{\bf r}_{yq})} \nonumber \\
 & & + I({\bf r}_{x \bar q},{\bf r}_{y \bar q}) - I({\bf r}_{xq},{\bf r}_{y \bar q}) - I({\bf r}_{x \bar q},{\bf r}_{yq}) \Bigr\}, \\
   \lefteqn{I({\bf r}_x,{\bf r}_y) = \kappa \frac{\pi}{2} \lambda^2 \; {\bf r}_y \cdot {\bf r}_x} \nonumber \\
 & & \int_0^1 \mbox{d}v \, \Biggl\{ \left(\frac{|v{\bf r}_y-{\bf r}_x|}{\lambda}\right)^2 K_2\left(\frac{|v{\bf r}_y-{\bf r}_x|}{\lambda}\right) \nonumber \\
& & \hspace{.95cm}+\left(\frac{|{\bf r}_y-v{\bf r}_x|}{\lambda}\right)^2 K_2\left(\frac{|{\bf r}_y-v{\bf r}_x|}{\lambda}\right)\Biggr\} \nonumber\\
& &+(1-\kappa)\pi\lambda^4 \left(\frac{|{\bf r}_y-{\bf r}_x|}{\lambda}\right)^3 K_3\left(\frac{|{\bf r}_y-{\bf r}_x|}{\lambda}\right), \nonumber \\
\end{eqnarray}
where $G_2$ is the gluon condensate, $\lambda=8a/3\pi$ is connected to the correlation length $a$ and $\kappa$ is a parameter which is related to the non-abelian character of the correlator. $K_{2,3}$ are the modified Bessel functions of second and third degree. The vectors ${\bf r}_{ij}$ with $i=x,y$ and $j=q,\bar q$ run from the coordinate origin to the positions of the quarks and antiquarks in transverse space.

%% file: results.tex
\begin{figure}[h]
  \unitlength1mm
    \begin{picture}(80,58)
      \put(-2,0){\epsfysize=57mm \epsffile{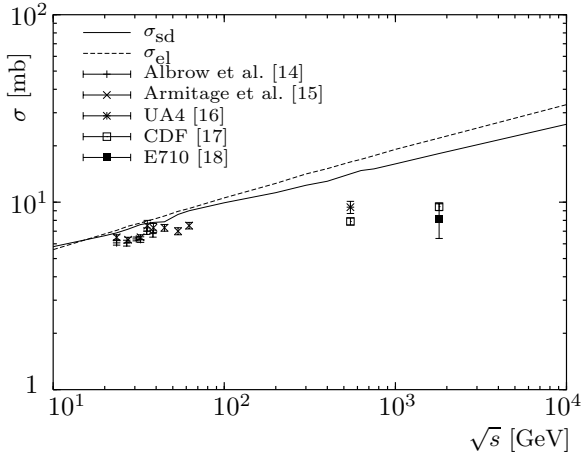}}
      \put(5,1){$10^1$}
      \put(27,1){$10^2$}
      \put(49,1){$10^3$}
      \put(71,1){$10^4$}
      \put(2,4){$1$}
      \put(0,28){$10^1$}
      \put(0,53){$10^2$}
      \small
      \put(61,-3){$\sqrt{s}$ [GeV]}
      \put(2,40){\begin{rotate}{90} $\sigma$ [mb]\end{rotate}}
      \scriptsize
      \put(16,43){$\begin{array}{l} \sigma_{\mbox{\scriptsize sd}} \\ \sigma_{\mbox{\scriptsize el}} \\ \mbox{Albrow et al.}~\cite{Albrow} \\ \mbox{Armitage et al.}~\cite{Armitage} \\ \mbox{UA4}~\cite{Bernard} \\ \mbox{CDF}~\cite{Abe} \\ \mbox{E710}~\cite{Amos} \end{array} $}
    \end{picture}
  \vspace*{-1.1cm}
  \caption{{\em The integrated single diffractive cross section as a function of $\sqrt{s}$}}\label{sigdiss_fig}
\vspace*{-.4cm}
\end{figure}

\section{Total and differential cross sections}\label{result_sect}

Now we consider diffractive proton-proton scattering, i.e. $h_1=h_2=p$. To calculate cross sections, we have to fix our free parameters, those of the MSV and the parameters of the proton wave function, i.e. the extension parameter $S_p$ and the width of the longitudinal momentum distribution $z_p$. The set of MSV parameters used in this work has been established in~\cite{DGKP}: $G_2=(501 \, \mathrm{MeV})^4, \, a=0.346 \, \mathrm{fm}, \, \kappa=0.74$.
The proton extension parameter is allowed to be energy dependent. From a fit of the total cross section as calculated in the model to the pomeron part of the Donnachie-Landshoff (DL) parametrisation for $\sigma_{\mbox{\scriptsize{tot}}}$~\cite{DL} we obtain the following connection between $S_p$ and $s$:
\begin{equation}\label{S2_chi2}
  S_p(s) = 0.624 \left( \frac{s}{\mbox{GeV}^2} \right)^{0.028} \mbox{ fm}.
\end{equation}
The width of the longitudinal momentum distribution has been chosen as $z_p=0.21$ which gives a best fit to the electric form factor of the proton calculated in the framework of our model~\cite{Diplom}.

\begin{figure}[h]
  \unitlength1mm
  \begin{center}
    \begin{picture}(80,58)
      \put(2,0){\epsfig{file=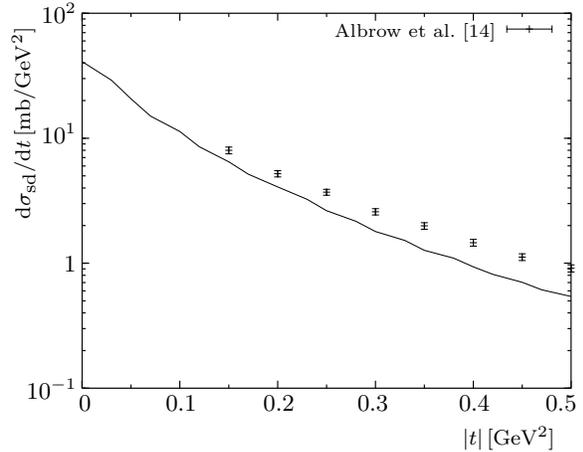,height=57mm,width=76mm}}
      \small
      \put(9,1.5){$0$}
      \put(20.5,1.5){$0.1$}
      \put(33.5,1.5){$0.2$}
      \put(46,1.5){$0.3$}
      \put(59.5,1.5){$0.4$}
      \put(71,1.5){$0.5$}
      \put(2,3){$10^{-1}$}
      \put(7,20){$1$}
      \put(4,37){$10^1$}
      \put(4,53){$10^2$}
      \footnotesize
      \put(60,-3){$|t| \, [\mbox{GeV}^2]$}
      \put(2.5,27){\begin{rotate}{90} $\mbox{d}\sigma_{\mbox{\scriptsize{sd}}} / \mbox{d}t \, [\mbox{mb}/\mbox{GeV}^2]$ \end{rotate}}
      \scriptsize
      \put(43,51.2){Albrow et al.~\cite{Albrow}}
    \end{picture}
  \end{center}
  \vspace*{-1cm}
  \caption{{\em The differential diffractive cross section} $\mbox{d}\sigma_{\mbox{\scriptsize{sd}}} / \mbox{d}t \, [\mbox{mb}/\mbox{GeV}^2]$ {\em at} $\sqrt{s}=23.5 \mbox{ GeV}$}\label{dsig_pp_diss_fig}
  \vspace*{-.9cm}
\end{figure}
With all parameters fixed, we can now do the numerical calculations for the scattering amplitude. We calculate the differential diffractive cross section from~(\ref{scat_amp}), where we insert~(\ref{chi2_result}) for the loop-loop correlation function and the wave and profile functions~(\ref{wave_funct}), (\ref{profile_funct}) in~(\ref{J_wave})
\begin{equation}
  \mbox{d}\sigma_{\mbox{\scriptsize{diff}}}=(2\pi)^4\frac{1}{2s} \left|\mathcal{T}_{fi}\right|^2 \mbox{d}^5\mathcal{P},
\end{equation}
where 
\begin{eqnarray}\label{P5}
  \mbox{d}^5\mathcal{P}=\frac{1}{(2\pi)^9} \frac{1}{4s \, z^\prime (1-z^\prime)} \mbox{d}^2P_{4T} \mbox{d}^2\Delta_{4T} \mbox{d}z^\prime
\end{eqnarray}
is the 5-dimensional phase space measure.

To obtain the cross section for single diffractive dissociation, we have to subtract the elastic contribution (cf.~\cite{Berger}) and then multiply by 2 to account for the reaction where the first proton breaks up. We then find for the integrated single diffractive cross section as a function of $\sqrt{s}$ the result shown in Fig.~\ref{sigdiss_fig}, where we have included our result for the integrated elastic cross section for comparison. Comparing our results to experimental data, one has to keep in mind that the overall normalisation uncertainty of the experiments is of $\mathcal{O}(10\%)$.

We now calculate the fraction $R=\sigma_{\mbox{\scriptsize{sd}}} / (\sigma_{\mbox{\scriptsize{el}}}+\sigma_{\mbox{\scriptsize{sd}}})$ of the diffractive dissociation cross section to the sum of the diffractive dissociation and the elastic parts and compare to experiment~\cite{PDG,Albrow,Armitage,Bernard,Abe,Amos}.
For c.m. energies $\sqrt{s}=$ 23.5, 546 and 1800 GeV we find $R=$ 0.49, 0.46 and 0.45, respectively, compared to the experimental values $R=0.49 \pm 0.01$, $0.41 \pm 0.02$ and $0.38 \pm 0.02$.
As can be seen from Fig.~\ref{sigdiss_fig} and the calculated $R$-values, our model is in qualitative agreement with the experimentally observed trend that the diffractive dissociation cross section grows more slowly with increasing energy than the elastic cross section.

The result for the differential cross section of the diffractive dissociation is shown in Fig.~\ref{dsig_pp_diss_fig}.
A fit to our result of the form $\mbox{d}\sigma_{\mbox{\scriptsize{sd}}} / \mbox{d}t = A \exp b \, t$ gives $b=7.3 \pm 0.2 \, \mbox{GeV}^{-2}$ compared to the experimental value $b=7.0 \pm 0.3 \, \mbox{GeV}^{-2}$~\cite{Albrow}, meaning that our model can reproduce the shape of the diffractive $|t|$-distribution reasonably well in the range of small $|t|$ considered here.

%% file: conclusion.tex
\section{Conclusion}\label{concl_sect}
In this work we have calculated integrated and differential cross sections for inelastic diffractive scattering at high c.m energies and small momentum transfer from the scattering amplitudes. In the ISR energy regime, our model describes the experimental data on proton-proton scattering well within the numerical and experimental errors and reproduces the experimentally observed features of diffractive scattering (e.g. the shrinkage of the ratio of the cross sections of diffractively excited to the sum of elastic scattering and diffractive excitation). Future investigations will cover the shapes of the diffractive mass spectra and the influence of higher terms in the cumulant expansion of the loop-loop correlation function.

A further field of application for our model will be the study of double diffractive dissociation reactions, where $C=-1$ contributions (odderon exchange) would be included already in our current approximation. Upcoming experiments at RHIC will be a rich source for new experimental data for both single and double diffractive reactions in hadronic reactions at high c.m. energies and therefore the study of inelastic diffractive scattering will remain an interesting and instructive subject for future studies.